\documentclass[preprint,superscriptaddress]{revtex4-1}
\usepackage{CJK} 
\usepackage{amsmath}
\usepackage{here}
\usepackage{mathptmx} %Use Times as default text font, and provide maths support
\usepackage[dvipdfmx]{graphicx}
\usepackage{bm} % bold math
\usepackage{color}
\usepackage{lipsum}
\date{\today}
\begin{document}
\begin{CJK*}{}{} 
\title{Selective imaging of the terahertz electric field of the phonon-polariton in LiNbO$ _3 $}
 \author{Keita~Matsumoto}
   \email{k-matsumoto@phys.kyushu-u.ac.jp}
 \affiliation{Department of Physics, Tokyo Institute of Technology, Tokyo 152-8551, Japan}
 \affiliation{Department of Physics, Kyushu University, Fukuoka 819-0385, Japan}
 \author{Takuya~Satoh}
  \email{satoh@phys.titech.ac.jp}
 \affiliation{Department of Physics, Tokyo Institute of Technology, Tokyo 152-8551, Japan}
\begin{abstract}
Coherent phonon-polaritons have attracted a considerable amount of interest owing to their relevance to nonlinear optics and terahertz (THz)-wave emissions. Therefore, it is important to analyze the THz electric-fields of phonon-polaritons. However, in the majority of previous measurements, only a single component of the THz electric field was detected. In this paper, we demonstrate that pump-probe electro-optical imaging measurements using the Stokes parameters of probe polarization enable the phase-resolved selective detection of THz electric-field components that are associated with the phonon-polariton. We experimentally distinguish the mode profiles of ordinary and extraordinary phonon-polaritons, and clarify the excitation mechanism as optical rectification. These results are explained by numerical calculations of Maxwell equations for the THz electric field. The technique of selectively observing the THz electric field components may be useful for designing efficient THz-wave emitters.
\end{abstract}
\maketitle
\end{CJK*}
\section{Introduction}
The phonon-polariton, which is a coupling mode of a terahertz (THz) electric field with a phonon in a medium \cite{Hopfield1958,Loudon1972,Feurer2003,Wahlstrand2003}, has been studied extensively and observed in non-centrosymmetric materials owing to its relevance to nonlinear optics and THz-wave emissions. The coherent phonon-polariton can be excited by an ultrashort optical pulse via impulsive stimulated Raman scattering (ISRS) \cite{Cavalleri2006,Wu2013,Ikegaya2015,Kuribayashi2018} or optical rectification \cite{Auston1984,Cheung1985,Hu1990,Stevens2001,Hebling2002,Suizu2009}, and the phase-matching condition results in Cherenkov radiation. In particular, this Cherenkov-type phonon-polariton has been used for efficient THz-wave emission and can be clearly observed in LiNbO$ _3 $ \cite{Crimmins2002,Stoyanov2002,Feurer2007,Yang2010,Wang2015,Ikegaya2015,Kuribayashi2018}, which is a typical ferroelectric material.

LiNbO$ _3 $ offers a wide range of optical applications, such as THz-wave emitters \cite{Kawase1996,Kawase2002,Yeh2007} and quasi-phase-matched devices \cite{Fejer1992,Yamada1993,Burns1994,Myers1995}, owing to its significant nonlinear optical effects and ferroelectricity. 
LiNbO$ _3 $ exhibits high optical-to-THz conversion efficiency, particularly when using the tilted-pump-pulse-front scheme \cite{Hebling2002,Stepanov2003,Hebling2004,Stepanov2005,Yeh2007,Hirori2011,Nagai2012} or Si-prism coupling method \cite{Kawase2001,Bodrov2008,Tani2011,Bakunov2012,Bakunov2014}. Thus, it is necessary to survey the response to the ultrashort pulse laser and to analyze the THz electric field. In the majority of previous pump-probe experiments on LiNbO$ _3 $, only one component of the THz electric field associated with the phonon-polariton was detected because a linearly polarized probe pulse was used. 

A recent experiment demonstrated that several dynamic magnetization components were selectively resolved with pump-probe measurements, by detecting the change in the Stokes parameters of a circularly polarized probe pulse \cite{Satoh2015}. The Stokes parameters ($S_0,\ S_1,\ S_2,\ S_3$) describe the complete polarization states of light, where $ S_0 $ denotes the light intensity, $S_1$ and $ S_2$ denote the light polarization direction, and $ S_3 $ represents the helicity of the circular polarization. Therefore, in addition to the magnetization, the detection of the change in the Stokes parameters $ S_1 $ and $ S_2 $ of the incident circularly polarized probe pulse can disclose the phase and amplitude of the THz electric field components selectively. 

In this study, we performed a detailed analysis of the THz electric field associated with the phonon-polariton in LiNbO$ _3 $ using the phase-resolved pump-probe imaging technique. The numerical calculations of the THz electric field along the $ x $ and $ y $ axes showed good agreement with the experimental detections of the Stokes parameters $ S_1 $ and $ S_2 $, respectively, indicating the capability of the selective detection of the THz electric field components. Moreover, we analyzed the dispersion relation and mode profiles of the excited phonon-polariton in terms of the Cherenkov radiation through optical rectification.

\section{Experimental Methods}
\begin{figure}[htbp]
	\centering
	\includegraphics[width=8.6cm]{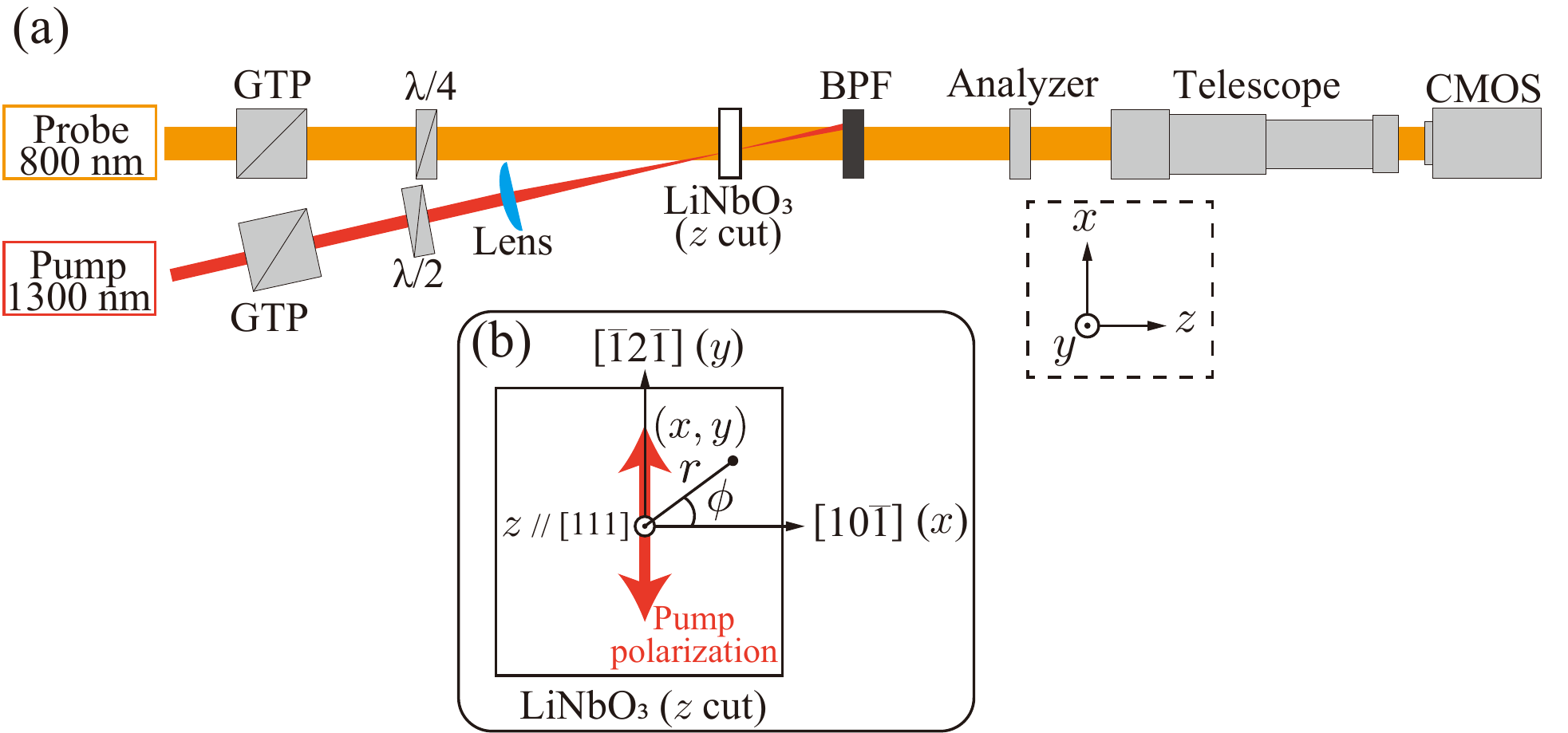}
	\caption{(a) Experimental configuration of pump-probe imaging. GTP: Glan--Taylor prism; $ \lambda/2 $: half-wave plate; $ \lambda/4 $: quarter-wave plate; BPF: band-pass filter, which passes a wavelength of 800 nm; CMOS: complementary metal oxide semiconductor camera. (b) Crystal orientation of LiNbO$ _3 $ ($ z $ cut), pump polarization, and definition of cylindrical coordinates.}
	\label{experimental_setup}
\end{figure}

Figure \ref{experimental_setup}(a) illustrates the experimental setup of the pump-probe electro-optical (EO) imaging measurement. The pump and probe pulses were generated by a Ti:sapphire regenerative amplifier with a repetition rate of 1 kHz and a pulse duration of $\tau= 60 $ fs. The optical parametric amplifier converted the linear polarized pump pulse to a pulse with the same pulse duration and 1300-nm central wavelength; the converted pulse was focused in a circular shape to with half width at half maximum of 35 $ \mu $m. The angle of incidence of the pump pulse was $6^\circ $. The probe pulse was circularly polarized with a central wavelength of 800 nm, incident perpendicularly on the sample surface without focusing, and was detected by a complementary metal oxide semiconductor camera \cite{Yoshimine2014}. For the pump and probe pulses, the pulse fluences were 160 mJ cm$ ^{-2} $ and 0.3 mJ cm$ ^{-2} $, respectively. All the measurements were performed at room temperature.

The pump polarization was oriented along the $ y $ axis, as illustrated in Fig. \ref{experimental_setup}(b), to minimize the influence of oblique incidence [see Fig. \ref{experimental_setup}(a)]. 
We measured the transmitted intensities for four settings of the analyzer, namely, $0^\circ$, $45^\circ$, $90^\circ$, and $135^\circ$ from the $ x $ axis. From these data, we obtained $ S_1$ by subtracting $ 90^\circ $ from $ 0^\circ $, and $ S_2 $ by subtracting $ 135^\circ $ from $ 45^\circ $, which are referred to as $ S_1 $ probe and $ S_2 $ probe, respectively.

Our sample was a uniaxial LiNbO$ _3 $ ($ z $-cut) single crystal with a point group of $ 3m $, grown by the Czochralski method and with a thickness of 500 $ \mu$m. The permittivity for the extraordinary wave polarized along the optic ($ z $) axis was $\varepsilon_\text{e}=26$, whereas the permittivity for the ordinary wave polarized along the $ x $ and $ y $ axes was $\varepsilon_\text{o}=41.5$ in the sub-THz regime \cite{Feurer2007}.
Thus, without pump excitation, no birefringence occurred in the $x$--$ y$ plane, enabling the transmission of a circularly polarized probe pulse. 

\section{Numerical calculations}
A pump pulse traveling in the $ z $ direction with a speed of $ c/n_\text{g} $ generates THz electric polarization $ {\mathbf P}^\text{THz} $ through optical rectification, where $ c$ is the speed of light in vacuum and $ n_\text{g}=2.27$ is the group refractive index at 1300 nm \cite{Zelmon1997}.
By using the optical rectification tensor [$ d_{yyy}=-8.8\times10^{-8} $, $ d_{zxx}=-12.4\times10^{-8}$ cm/statV ($ d_{yyy}=-37$ and $ d_{zxx}=-52$ pm/V in SI units)] for LiNbO$ _3 $ \cite{Bakunov2014}, the THz electric polarization in a cylindrical coordinate $ (r,\phi,z)$, as indicated in Fig. \ref{experimental_setup}(b) is expressed as follows:
\begin{align}
{\mathbf P}^\text{THz}&=(P^\text{THz}_r,P^\text{THz}_\phi,P^\text{THz}_z)\cdot G(r){F}(t)\nonumber\\
&=(d_{yyy}\sin\phi,d_{yyy}\cos\phi,d_{zxx})I\cdot G(r){F}(t),
\end{align}
where $ I $ is the pump-pulse intensity, $ G(r) = \exp[-r^2/(2l_{\perp}^2)] $, $l_\perp=25.5\ \mu$m was set for our calculation, and ${F}(t)=\exp[-(t-zn_\text{g}/c)^2/\tau^2]$.
We note that excitation of the phonon-polariton via ISRS was not likely to occur because the pump polarization contained only the ordinary wave \cite{Loudon1964,Ikegaya2015,Kuribayashi2018}, as discussed in the next section.

We numerically calculated the THz electric field $ \mathbf{E}^\text{THz} $ of the phonon-polariton to explain the experimental results. In the calculation, we referred to a theoretical framework \cite{Bakunov2005,Bodrov2008,Bakunov2012,Bakunov2014} that describes the one-dimensional propagation of the THz electric field in LiNbO$ _3 $ ($ x $ cut) by using the Maxwell equations. We extended this theory to two-dimensional propagation in LiNbO$ _3 $ ($ z $ cut) with the cylindrical coordinate system:
\begin{align}
\nabla\times\nabla\times\tilde{\mathbf E}^\text{THz}=
\frac{\omega^2}{c^2}
\begin{pmatrix}
\varepsilon_\text{o}&0&0\\
0&\varepsilon_\text{o}&0\\
0&0&\varepsilon_\text{e}
\end{pmatrix}
\tilde{\mathbf E}^\text{THz}+\frac{4\pi\omega^2}{c^2}\tilde{\mathbf P}^\text{THz},\label{maxewll_eqn_theory}
\end{align}
where $ \nabla=(\partial/\partial r,0,-i\omega n_\text{g}/c) $ and $\ \tilde{}\ $ represents quantities in the Fourier space ($ \omega $). The influence of the phonon is incorporated into $ \varepsilon_\text{o} $ and $ \varepsilon_\text{e} $, which we assumed to be non-dispersive, and the validity thereof is discussed in the following section. 

In this calculation, we neglected the propagation along the angular direction $ \phi$ because the pump spot was focused circularly. Thus, according to Eq. \eqref{maxewll_eqn_theory}, the governing wave equations can be expressed as follows:

\begin{align}
&\dfrac{\partial^2 \tilde{E}_{r}^\text{THz}}{\partial {r}^2}+\dfrac{\omega^2}{c^2}\varepsilon_\text{eff} \tilde{E}_{r}^\text{THz}\nonumber\\
&\quad=\dfrac{4\pi}{\varepsilon_\text{o}}\left[-\left(\dfrac{\partial^2G(r)}{\partial r^2}+\varepsilon_\text{e}\dfrac{\omega^2}{c^2}G(r)\right)P_r+i\dfrac{\omega}{c}n_\text{g}\dfrac{\partial G(r)}{\partial r}P_z\right]\tilde{F}(\omega)\nonumber\\\vspace{0.1cm}
&\dfrac{\partial^2 \tilde{E}_{\phi}^\text{THz}}{\partial {r}^2}+\dfrac{\omega^2}{c^2}\dfrac{\varepsilon_\text{o}}{\varepsilon_\text{e}}\varepsilon_\text{eff} \tilde{E}_{\phi}^\text{THz}\nonumber\\
&\quad= -\dfrac{4\pi\omega^2}{c^2}P_\phi G({r})\tilde{F}(\omega)\nonumber\\
&\dfrac{\partial^2 \tilde{E}_z^\text{THz}}{\partial {r}^2}+\dfrac{\omega^2}{c^2}\varepsilon_\text{eff} \tilde{E}_z^\text{THz} \nonumber\\
&\quad= \dfrac{4\pi i\omega n_\text{g}}{\varepsilon_\text{o} c}\left[\dfrac{\partial G({r})}{\partial {r}}P_r+i\dfrac{\omega\varepsilon_\text{o}}{cn_\text{g}\varepsilon_\text{e}}\varepsilon_\text{eff}G({r})P_z\right]\tilde{F}(\omega),\label{governing_equations}
\end{align}

where $\varepsilon_\text{eff}=\varepsilon_\text{e}(1-n_\text{g}^2/\varepsilon_\text{o})$ is the effective permittivity. Finally, we used the inverse Fourier transform of Eq. \eqref{governing_equations} and applied a coordinate transformation from the cylindrical to Cartesian coordinates, so as to obtain the spatial dynamics of $ {\mathbf E}^\text{THz}=(E_x^\text{THz},E_y^\text{THz},E_z^\text{THz})$.

\section{Results and discussion}
\begin{figure}[htbp]
	\centering
	\includegraphics[width=13.6cm]{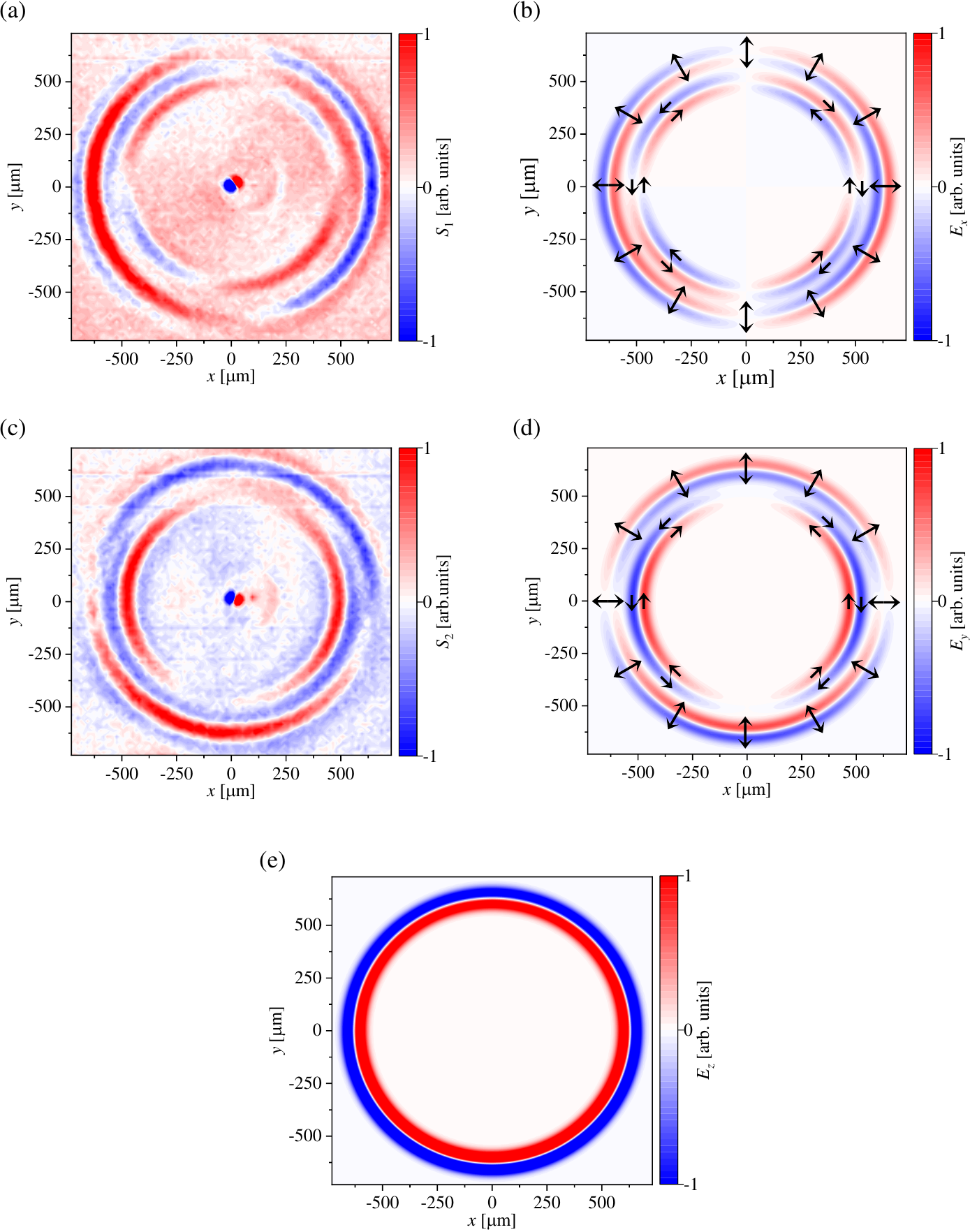}
	\caption{(a) Experimental result of $ S_1$-probe configuration and (b) corresponding numerical calculation of $ E_x^\text{THz} $; (c) experimental result of $S_2$-probe configuration and (d) corresponding calculation of $ E_y^\text{THz}$ at 10 ps after pump excitation. The black arrows in (b) and (d) indicate the directions of the THz electric field. (e) Calculated waveform of $ E_z^\text{THz} $.}
	\label{Results}
\end{figure}
\begin{figure}[t]
	\centering
	\includegraphics[width=14cm]{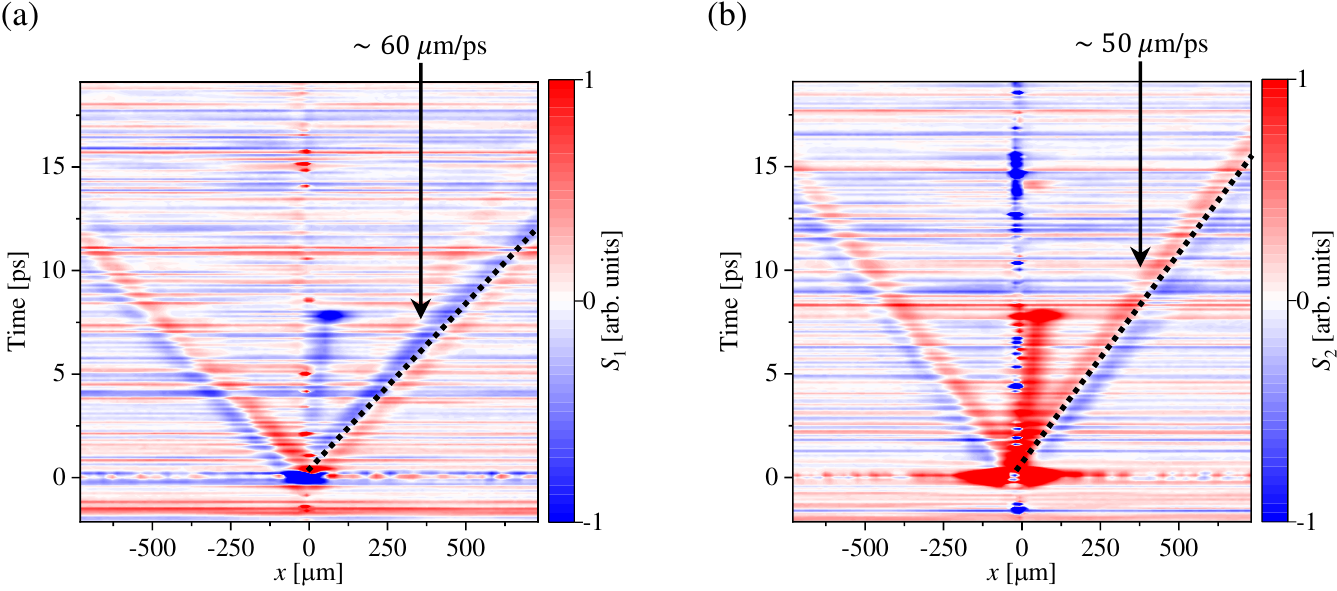}
	\caption{Experimentally observed spatiotemporal map along the $ x $ axis for (a) $ S_1 $- and (b) $ S_2 $-probe configurations. The dashed lines on each panel express the propagation speed.}
	\label{Results_spatiotemporal}
\end{figure}
Once the THz electric field $ \mathbf{E}^\text{THz}$ is generated by the pump pulse, birefringence is induced in the $ x$--$y$ plane, which modulates the polarization of the circularly polarized probe pulse via the EO effect \cite{Tani2011}. As the EO effect yields the permittivity change, the inverse matrix of the in-plane permittivity $(\varepsilon^{-1})_{ij}$ is expressed as follows:
\begin{widetext}
\begin{align}
(\varepsilon^{-1})_{ij}=\delta_{ij}/\varepsilon_\text{o}+\sum_{k}r_{ijk}E_{k}^\text{THz}
=\begin{pmatrix}
1/\varepsilon_\text{o}+r_{xzx}E_z^\text{THz}-r_{yyy}E_y^\text{THz}&-r_{yyy}E_x^\text{THz}\\
-r_{yyy}E_x^\text{THz}&1/\varepsilon_\text{o}+r_{xzx}E_z^\text{THz}+r_{yyy}E_y^\text{THz}
\end{pmatrix}
,
\end{align}
\end{widetext}
where $ \delta_{ij}$ is the Kronecker delta, $ r_{ijk} $ is the EO tensor \cite{Shen_nonlinearOptics,Feurer2007} for a $ 3m $ crystal, $\{i,j\}=\{ x,y\}$, and $ k=\{x,y,z\} $. Because the EO modulation is sufficiently small, we can approximately determine the permittivity change $ \delta\varepsilon_{ij} $ from the permittivity $ \varepsilon_{ij}=\varepsilon_\text{o}\delta_{ij}+\delta\varepsilon_{ij} $, as follows:
\begin{align}
\delta\varepsilon_{ij}\approx
\varepsilon_\text{o}^2\begin{pmatrix}
-r_{xzx}E_z^\text{THz}+r_{yyy}E_y^\text{THz}&r_{yyy}E_x^\text{THz}\\
r_{yyy}E_x^\text{THz}&-r_{xzx}E_z^\text{THz}-r_{yyy}E_y^\text{THz}\\
\end{pmatrix}.
\end{align}

Previous research \cite{Satoh2015,Khan2020} demonstrated the relation between the permittivity change and Stokes parameters: $ S_1\propto\delta\varepsilon_{xy}+\delta\varepsilon_{yx}$ and $ S_2\propto\delta\varepsilon_{xx}-\delta\varepsilon_{yy} $. Therefore, in our setup, $ S_1\propto E_x^\text{THz} $ and $ S_2\propto E_y^\text{THz} $, which implied selective detection of the THz electric field components by using the Stokes parameters.

Figures \ref{Results}(a) and \ref{Results}(b) depict the comparison of the experimental results by the $ S_1 $-probe configuration with the calculated waveform of $ E_x^\text{THz} $ at 10 ps following pump excitation, and the similarity of these results confirmed that $S_1\propto E_x^\text{THz}$. At $ x=y=0 $, the pump pulse was shone on the sample at $ t=0 $ ps. Thereafter, the THz electric polarization and electric field were generated through optical rectification, such that the THz electric field of the phonon-polariton was propagated in the $ r $ direction with two distinct speeds of $ \approx 50 $ and $ 60\ \mu$m/ps. Similarly, Figs. \ref{Results}(c) and \ref{Results}(d) depict the comparison results of the experiment by $ S_2 $-probe configuration and the numerical calculation of $ E_y^\text{THz} $ at 10 ps following pump excitation. The propagating speeds of both wave forms were identical to those in Fig. \ref{Results}(a), and the relation of $S_2\propto E_y^\text{THz}$ was confirmed. The agreements between the experiments and numerical calculations prove the selective detection of the THz electric-field components. Furthermore, the calculated wave form of $ E_z^\text{THz} $ is illustrated in Fig. \ref{Results}(e) (see also the movie in the Supplemental Material \cite{SM20}). The wave form of $ E_z^\text{THz} $ was non-zero only for the outer wave form, which implied that the THz electric field of the inner wave form did not contain an extraordinary wave.

The black arrows mapped onto Figs. \ref{Results}(b) and \ref{Results}(d) indicate the direction of the THz electric field. The THz electric field of the inner wave form was polarized along $\phi$ and perpendicular to the propagation direction, whereas the outer wave-form polarization was along $ r $ and $ z $. Therefore, these propagating modes could be attributed to the ordinary and extraordinary phonon-polaritons \cite{Loudon1964,Claus1972,Delbart1998} for the inner and outer wave forms, respectively. 

Figures \ref{Results_spatiotemporal}(a) and \ref{Results_spatiotemporal}(b) illustrate the spatiotemporal maps of the wave form propagating along the $ x $ axis, which were obtained experimentally in the $ S_1 $- and $ S_2 $-probe configurations, respectively. The spatial profiles along the $ y $ axis were averaged; thus, Fig. \ref{Results_spatiotemporal}(a) reflects the extraordinary phonon-polariton propagating at a speed of $ \approx60 \ \mu $m/ps, whereas Fig. \ref{Results_spatiotemporal}(b) indicates the ordinary phonon-polariton propagating at a speed of $ \approx 50\ \mu $m/ps. These speeds were determined by Eq. \eqref{governing_equations}: $ E_r^\text{THz}$ and $E_z^\text{THz}$ had a wave number of $ \omega\sqrt{\varepsilon_\text{eff}}/c $, with a corresponding speed of $c/\sqrt{\varepsilon_\text{eff}}\approx60\ \mu$m/ps; $ E_\phi^\text{THz} $ had a wave number of $ \omega\sqrt{\varepsilon_\text{o}\varepsilon_\text{eff}/\varepsilon_\text{e}}/c$, with a corresponding speed of $c/\sqrt{\varepsilon_\text{eff}\varepsilon_\text{e}/\varepsilon_\text{o}}\approx50\ \mu$m/ps.

\begin{figure}[htbp]
	\centering
	\includegraphics[width=8.6cm]{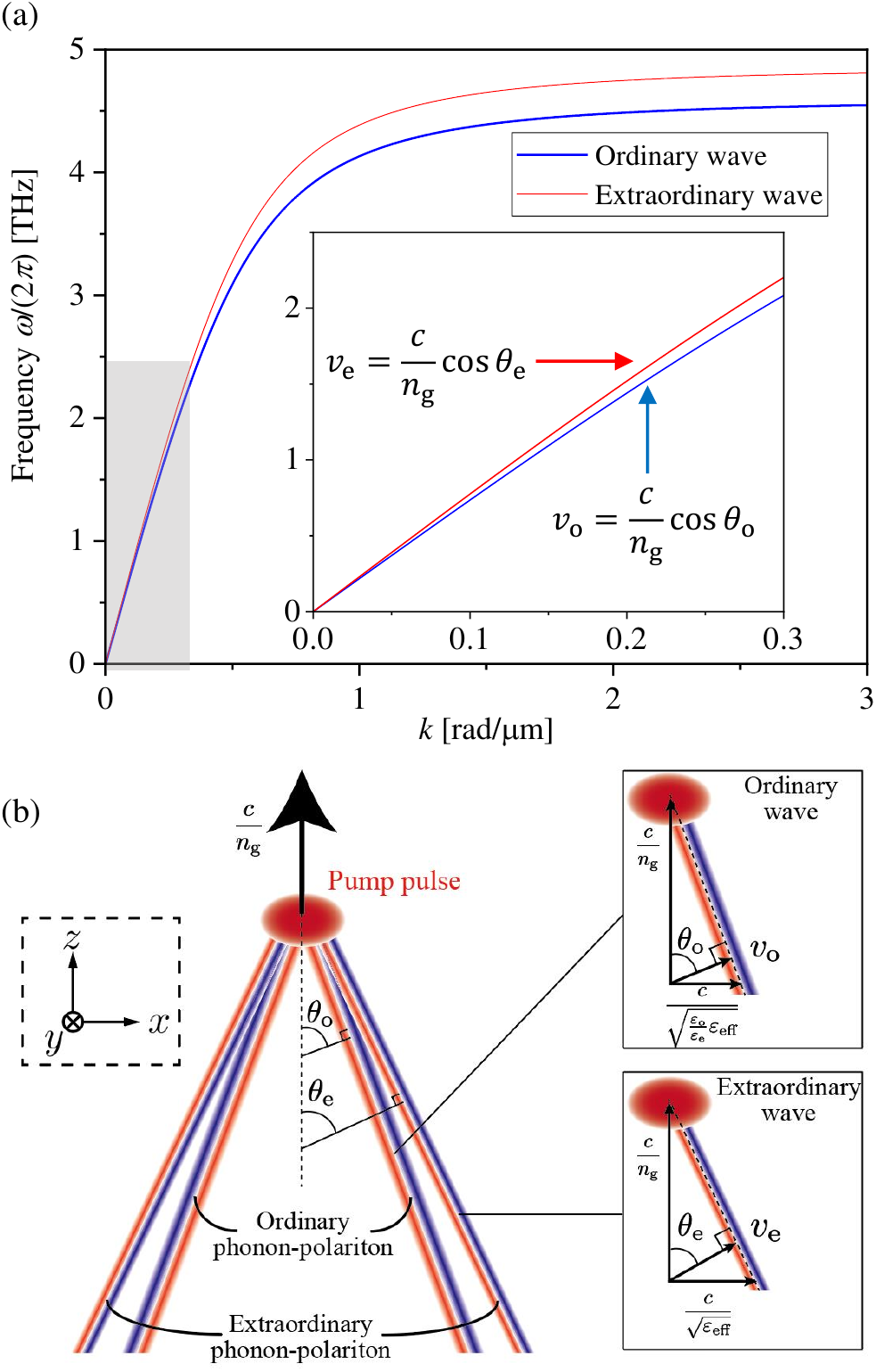}
	\caption{(a) Dispersion relation of phonon-polariton. The inset presents a magnification of the shaded region, indicating the different propagation speeds of the ordinary and extraordinary phonon-polaritons. The horizontal axis is a wave number perpendicular to the wave front of the phonon-polariton. (b) Schematic of ordinary and extraordinary phonon-polariton modes propagating at Cherenkov angles of $ \theta_\text{o} $ and $ \theta_\text{e} $. The propagation directions of the wave fronts are illustrated in the right panels.}
	\label{cherenkov_explanation}
\end{figure}

The theoretical dispersion curves of these phonon-polariton modes are depicted in Fig. \ref{cherenkov_explanation}(a). The blue curve is the ordinary one expressed by 
\begin{align}
\frac{k^2c^2}{\omega^2}=\frac{\omega_\text{o}^2\varepsilon_\text{o}-\omega^2\varepsilon_\text{o}^\infty}{\omega_\text{o}^2-\omega^2},\label{dispersion_ordinary}
\end{align}
where $ k $ is a wave number perpendicular to the wave front of the phonon-polariton, $ \omega_\text{o}/(2\pi) = 4.6$ THz is the frequency of the $ E $ phonon at 152 cm$ ^{-1} $, polarized perpendicular to the optic axis \cite{Barker1967}, and $ \varepsilon_\text{o}^\infty=19.5$ is the permittivity of the ordinary wave at the limit of $ \omega\rightarrow\infty$ \cite{Feurer2007}. 

The red curve indicates the extraordinary phonon-polariton, represented by
\begin{align}
\frac{k^2c^2}{\omega^2}=\frac{\left[\frac{\omega_\text{e}^2\varepsilon_\text{e}-\omega^2\varepsilon_\text{e}^\infty}{\omega_\text{e}^2-\omega^2}\right]\left[\frac{\omega_\text{o}^2\varepsilon_\text{o}-\omega^2\varepsilon_\text{o}^\infty}{\omega_\text{o}^2-\omega^2}\right]}{\left[\frac{\omega_\text{e}^2\varepsilon_\text{e}-\omega^2\varepsilon_\text{e}^\infty}{\omega_\text{e}^2-\omega^2}\right]\cos^2\theta_\text{e}+\left[\frac{\omega_\text{o}^2\varepsilon_\text{o}-\omega^2\varepsilon_\text{o}^\infty}{\omega_\text{o}^2-\omega^2}\right]\sin^2\theta_\text{e}},\label{dispersion_extraordinary}
\end{align}
where $ \theta_\text{e} $ is the propagation angle from the $ z $ axis, $ \omega_\text{e}/(2\pi)=7.4 $ THz is the frequency of the $ E $ phonon at 248 cm$ ^{-1} $, polarized parallel to the optic axis \cite{Barker1967}, and $ \varepsilon_\text{e}^\infty=10$ is the permittivity of the extraordinary wave at the limit of $ \omega\rightarrow\infty $ \cite{Feurer2007}. 

As the spatial distribution of the pump pulse is $ \exp[-r^2/(2l_\perp^2)]$ according to Eq. \eqref{governing_equations}, the wave number of the excited phonon-polariton follows $\propto \exp(-k^2l_\perp^2/2)$ \cite{Satoh2012,Yoshimine2017,Tilburg2017,Kmatsumoto2020}. Thus, the upper limit of the excited wave number is $ 2\pi/l_\perp\approx0.25 $ rad$ /\mu $m. At such a small wave number, the dispersions are approximately linear with slopes of ${v}_\text{o}=\omega/k=c/\sqrt{\varepsilon_\text{o}}=c\cos\theta_\text{o}/n_\text{g} $ and $v_\text{e}=\omega/k=c/\sqrt{n_\text{g}^2+\varepsilon_\text{eff}}=c\cos\theta_\text{e}/n_\text{g} $ for the ordinary and extraordinary photon-polaritons, respectively, where $ \theta_\text{o} $ and $ \theta_\text{e} $ are the Cherenkov angles, as schematically depicted in Fig. \ref{cherenkov_explanation}(b). 

The group velocity of the pump pulse $c/n_\text{g}$ is faster than the phase velocity of the generated THz electric field, leading to Cherenkov radiation with angles of $ \theta_\text{o}$ and $ \theta_\text{e} $ for the ordinary and extraordinary phonon-polaritons, respectively. These angles can be obtained from equations using the variables in the panels on the right side of Fig. \ref{cherenkov_explanation}(b), as follows:
\begin{align}
\theta_\text{o}&=\tan^{-1}\left(\sqrt{\frac{\varepsilon_\text{o}}{\varepsilon_\text{e}}}\frac{\sqrt{\varepsilon_\text{eff}}}{n_\text{g}}\right)=\tan^{-1}\left(\sqrt{\frac{\varepsilon_\text{o}-n_\text{g}^2}{n_\text{g}^2}}\right)=64.5^\circ\\
\theta_\text{e}&=\tan^{-1}\left(\frac{\sqrt{\varepsilon_\text{eff}}}{n_\text{g}}\right)=\tan^{-1}\left[\sqrt{\frac{\varepsilon_\text{e}}{n_\text{g}^2}\left(1-\frac{n_\text{g}^2}{\varepsilon_\text{o}}\right)}\right]=69.3^\circ.
\end{align}

Finally, we discuss the possible excitation of the phonon-polariton through ISRS.
The Raman tensors of the $ E $ phonon in the $ x$--$y $ plane are \cite{Loudon1964,Khan2020}
\begin{align}
\hat{E}(-x)=
\begin{pmatrix}
0&-f\\
-f&0
\end{pmatrix}
\quad\text{and}\quad \hat{E}(y)=
\begin{pmatrix}
f&0\\
0&-f
\end{pmatrix},
\end{align}
where $ f $ is a component of the Raman tensor. The driving force of the $ E $ phonon is obtained by operating the pump polarization vector on the tensors as
\begin{align}
F_{E(x)}&=I\cdot F(t)
\begin{pmatrix}
0&1
\end{pmatrix}
\begin{pmatrix}
\ f&0\\
0&-f
\end{pmatrix}
\begin{pmatrix}
0\\1
\end{pmatrix}
=-I\cdot F(t)c
,\nonumber\\
F_{E(y)}&=I\cdot F(t)
\begin{pmatrix}
0&1
\end{pmatrix}
\begin{pmatrix}
0&-f\\
-f&0
\end{pmatrix}
\begin{pmatrix}
0\\1
\end{pmatrix}
=0.\label{driving_force_Ex_phonon}
\end{align}
Thus, the permittivity modulation $\delta\varepsilon^\text{ISRS}_{ij}$ is
\begin{align}
\delta\varepsilon^\text{ISRS}_{ij}\propto
F_{E(x)}\hat{E}(x)+F_{E(y)}\hat{E}(y)=
I\cdot F(t)
\begin{pmatrix}
-f^2&0\\
0&\ f^2
\end{pmatrix}.\label{permittivit_modulation_ISRS_Ephonon}
\end{align}
Equation \eqref{permittivit_modulation_ISRS_Ephonon} indicates the signal of $ S_2 $ probe because $ \delta\varepsilon^\text{ISRS}_{xx}-\delta\varepsilon^\text{ISRS}_{yy}$ is finite, while no signal can be expected for the $ S_1 $ probe because $ \delta\varepsilon^\text{ISRS}_{xy}+\delta\varepsilon^\text{ISRS}_{yx}$ is zero. 
However, as depicted in Figs. \ref{Results} and \ref{Results_spatiotemporal}, the signals in the $ S_1 $ probe were discernible, leading us to conclude that the excitation was mainly caused by optical rectification rather than ISRS.

\section{Conclusions}
We demonstrated a selective detection technique of the electric-field components of a THz wave propagating as a phonon-polariton in LiNbO$ _3 $ ($ z $ cut) by observing the Stokes parameters $ S_1 $ and $ S_2 $, which can provide more information regarding the phonon-polariton than previous studies. The experimental results were confirmed by the numerical calculations of $ E_x^\text{THz} $ and $ E_y^\text{THz} $ using the Maxwell equations. This technique enables phonon-polaritons with two different mode profiles of the THz electric field to be distinguished: the ordinary phonon-polariton that is polarized perpendicular to both the propagation direction and optic axis, and the extraordinary phonon-polariton. Based on the experimental results, we can conclude that the excitation mechanism is optical rectification rather than ISRS. The detection technique may provide a tool for understanding the fundamental nature of the propagation of THz waves in anisotropic materials.

\section*{Acknowledgement}
This study was supported by the Japan Society for the Promotion of Science (JSPS) KAKENHI (Grants No. JP19H01828, No. JP19H05618, No. JP19J21797, No. JP19K21854, and No. JP26103004) and the JSPS Core-to-Core Program (A. Advanced Research Networks). K.M. would like to thank the Research Fellowship for Young Scientists by the JSPS.
\bibliographystyle{apsrev4-1}
\bibliography{mc}

%merlin.mbs apsrev4-1.bst 2010-07-25 4.21a (PWD, AO, DPC) hacked
%Control: key (0)
%Control: author (72) initials jnrlst
%Control: editor formatted (1) identically to author
%Control: production of article title (-1) disabled
%Control: page (0) single
%Control: year (1) truncated
%Control: production of eprint (0) enabled
\begin{thebibliography}{51}%
\makeatletter
\providecommand \@ifxundefined [1]{%
 \@ifx{#1\undefined}
}%
\providecommand \@ifnum [1]{%
 \ifnum #1\expandafter \@firstoftwo
 \else \expandafter \@secondoftwo
 \fi
}%
\providecommand \@ifx [1]{%
 \ifx #1\expandafter \@firstoftwo
 \else \expandafter \@secondoftwo
 \fi
}%
\providecommand \natexlab [1]{#1}%
\providecommand \enquote  [1]{``#1''}%
\providecommand \bibnamefont  [1]{#1}%
\providecommand \bibfnamefont [1]{#1}%
\providecommand \citenamefont [1]{#1}%
\providecommand \href@noop [0]{\@secondoftwo}%
\providecommand \href [0]{\begingroup \@sanitize@url \@href}%
\providecommand \@href[1]{\@@startlink{#1}\@@href}%
\providecommand \@@href[1]{\endgroup#1\@@endlink}%
\providecommand \@sanitize@url [0]{\catcode `\\12\catcode `\$12\catcode
  `\&12\catcode `\#12\catcode `\^12\catcode `\_12\catcode `\%12\relax}%
\providecommand \@@startlink[1]{}%
\providecommand \@@endlink[0]{}%
\providecommand \url  [0]{\begingroup\@sanitize@url \@url }%
\providecommand \@url [1]{\endgroup\@href {#1}{\urlprefix }}%
\providecommand \urlprefix  [0]{URL }%
\providecommand \Eprint [0]{\href }%
\providecommand \doibase [0]{http://dx.doi.org/}%
\providecommand \selectlanguage [0]{\@gobble}%
\providecommand \bibinfo  [0]{\@secondoftwo}%
\providecommand \bibfield  [0]{\@secondoftwo}%
\providecommand \translation [1]{[#1]}%
\providecommand \BibitemOpen [0]{}%
\providecommand \bibitemStop [0]{}%
\providecommand \bibitemNoStop [0]{.\EOS\space}%
\providecommand \EOS [0]{\spacefactor3000\relax}%
\providecommand \BibitemShut  [1]{\csname bibitem#1\endcsname}%
\let\auto@bib@innerbib\@empty
%</preamble>
\bibitem [{\citenamefont {Hopfield}(1958)}]{Hopfield1958}%
  \BibitemOpen
  \bibfield  {author} {\bibinfo {author} {\bibfnamefont {J.~J.}\ \bibnamefont
  {Hopfield}},\ }\href@noop {} {\bibfield  {journal} {\bibinfo  {journal}
  {Phys. Rev.}\ }\textbf {\bibinfo {volume} {112}},\ \bibinfo {pages} {1555}
  (\bibinfo {year} {1958})}\BibitemShut {NoStop}%
\bibitem [{\citenamefont {Barker}\ and\ \citenamefont
  {Loudon}(1972)}]{Loudon1972}%
  \BibitemOpen
  \bibfield  {author} {\bibinfo {author} {\bibfnamefont {A.~S.}\ \bibnamefont
  {Barker}}\ and\ \bibinfo {author} {\bibfnamefont {R.}~\bibnamefont
  {Loudon}},\ }\href@noop {} {\bibfield  {journal} {\bibinfo  {journal} {Rev.
  Mod. Phys.}\ }\textbf {\bibinfo {volume} {44}},\ \bibinfo {pages} {18}
  (\bibinfo {year} {1972})}\BibitemShut {NoStop}%
\bibitem [{\citenamefont {Feurer}\ \emph {et~al.}(2003)\citenamefont {Feurer},
  \citenamefont {Vaughan},\ and\ \citenamefont {Nelson}}]{Feurer2003}%
  \BibitemOpen
  \bibfield  {author} {\bibinfo {author} {\bibfnamefont {T.}~\bibnamefont
  {Feurer}}, \bibinfo {author} {\bibfnamefont {J.~C.}\ \bibnamefont {Vaughan}},
  \ and\ \bibinfo {author} {\bibfnamefont {K.~A.}\ \bibnamefont {Nelson}},\
  }\href {\doibase 10.1126/science.1078726} {\bibfield  {journal} {\bibinfo
  {journal} {Science}\ }\textbf {\bibinfo {volume} {299}},\ \bibinfo {pages}
  {374} (\bibinfo {year} {2003})}\BibitemShut {NoStop}%
\bibitem [{\citenamefont {Wahlstrand}\ and\ \citenamefont
  {Merlin}(2003)}]{Wahlstrand2003}%
  \BibitemOpen
  \bibfield  {author} {\bibinfo {author} {\bibfnamefont {J.~K.}\ \bibnamefont
  {Wahlstrand}}\ and\ \bibinfo {author} {\bibfnamefont {R.}~\bibnamefont
  {Merlin}},\ }\href@noop {} {\bibfield  {journal} {\bibinfo  {journal} {Phys.
  Rev. B}\ }\textbf {\bibinfo {volume} {68}},\ \bibinfo {pages} {054301}
  (\bibinfo {year} {2003})}\BibitemShut {NoStop}%
\bibitem [{\citenamefont {Cavalleri}\ \emph {et~al.}(2006)\citenamefont
  {Cavalleri}, \citenamefont {Wall}, \citenamefont {Simpson}, \citenamefont
  {Statz}, \citenamefont {Ward}, \citenamefont {Nelson}, \citenamefont {Rini},\
  and\ \citenamefont {Schoenlein}}]{Cavalleri2006}%
  \BibitemOpen
  \bibfield  {author} {\bibinfo {author} {\bibfnamefont {A.}~\bibnamefont
  {Cavalleri}}, \bibinfo {author} {\bibfnamefont {S.}~\bibnamefont {Wall}},
  \bibinfo {author} {\bibfnamefont {C.}~\bibnamefont {Simpson}}, \bibinfo
  {author} {\bibfnamefont {E.}~\bibnamefont {Statz}}, \bibinfo {author}
  {\bibfnamefont {D.~W.}\ \bibnamefont {Ward}}, \bibinfo {author}
  {\bibfnamefont {K.~A.}\ \bibnamefont {Nelson}}, \bibinfo {author}
  {\bibfnamefont {M.}~\bibnamefont {Rini}}, \ and\ \bibinfo {author}
  {\bibfnamefont {R.~W.}\ \bibnamefont {Schoenlein}},\ }\href@noop {}
  {\bibfield  {journal} {\bibinfo  {journal} {Nature (London)}\ }\textbf
  {\bibinfo {volume} {442}},\ \bibinfo {pages} {664} (\bibinfo {year}
  {2006})}\BibitemShut {NoStop}%
\bibitem [{\citenamefont {Wu}\ \emph {et~al.}(2013)\citenamefont {Wu},
  \citenamefont {Chen}, \citenamefont {Zhang},\ and\ \citenamefont
  {Xu}}]{Wu2013}%
  \BibitemOpen
  \bibfield  {author} {\bibinfo {author} {\bibfnamefont {Q.}~\bibnamefont
  {Wu}}, \bibinfo {author} {\bibfnamefont {Q.-Q.}\ \bibnamefont {Chen}},
  \bibinfo {author} {\bibfnamefont {B.}~\bibnamefont {Zhang}}, \ and\ \bibinfo
  {author} {\bibfnamefont {J.-J.}\ \bibnamefont {Xu}},\ }\href@noop {}
  {\bibfield  {journal} {\bibinfo  {journal} {Front. Phys.}\ }\textbf {\bibinfo
  {volume} {8}},\ \bibinfo {pages} {217} (\bibinfo {year} {2013})}\BibitemShut
  {NoStop}%
\bibitem [{\citenamefont {Ikegaya}\ \emph {et~al.}(2015)\citenamefont
  {Ikegaya}, \citenamefont {Sakaibara}, \citenamefont {Minami}, \citenamefont
  {Katayama},\ and\ \citenamefont {Takeda}}]{Ikegaya2015}%
  \BibitemOpen
  \bibfield  {author} {\bibinfo {author} {\bibfnamefont {Y.}~\bibnamefont
  {Ikegaya}}, \bibinfo {author} {\bibfnamefont {H.}~\bibnamefont {Sakaibara}},
  \bibinfo {author} {\bibfnamefont {Y.}~\bibnamefont {Minami}}, \bibinfo
  {author} {\bibfnamefont {I.}~\bibnamefont {Katayama}}, \ and\ \bibinfo
  {author} {\bibfnamefont {J.}~\bibnamefont {Takeda}},\ }\href {\doibase
  10.1063/1.4928480} {\bibfield  {journal} {\bibinfo  {journal} {Appl. Phys.
  Lett.}\ }\textbf {\bibinfo {volume} {107}},\ \bibinfo {pages} {062901}
  (\bibinfo {year} {2015})}\BibitemShut {NoStop}%
\bibitem [{\citenamefont {Kuribayashi}\ \emph {et~al.}(2018)\citenamefont
  {Kuribayashi}, \citenamefont {Motoyama}, \citenamefont {Arashida},
  \citenamefont {Katayama},\ and\ \citenamefont {Takeda}}]{Kuribayashi2018}%
  \BibitemOpen
  \bibfield  {author} {\bibinfo {author} {\bibfnamefont {T.}~\bibnamefont
  {Kuribayashi}}, \bibinfo {author} {\bibfnamefont {T.}~\bibnamefont
  {Motoyama}}, \bibinfo {author} {\bibfnamefont {Y.}~\bibnamefont {Arashida}},
  \bibinfo {author} {\bibfnamefont {I.}~\bibnamefont {Katayama}}, \ and\
  \bibinfo {author} {\bibfnamefont {J.}~\bibnamefont {Takeda}},\ }\href
  {\doibase 10.1063/1.5021379} {\bibfield  {journal} {\bibinfo  {journal} {J.
  Appl. Phys.}\ }\textbf {\bibinfo {volume} {123}},\ \bibinfo {pages} {174103}
  (\bibinfo {year} {2018})}\BibitemShut {NoStop}%
\bibitem [{\citenamefont {Auston}\ \emph {et~al.}(1984)\citenamefont {Auston},
  \citenamefont {Cheung}, \citenamefont {Valdmanis},\ and\ \citenamefont
  {Kleinman}}]{Auston1984}%
  \BibitemOpen
  \bibfield  {author} {\bibinfo {author} {\bibfnamefont {D.~H.}\ \bibnamefont
  {Auston}}, \bibinfo {author} {\bibfnamefont {K.~P.}\ \bibnamefont {Cheung}},
  \bibinfo {author} {\bibfnamefont {J.~A.}\ \bibnamefont {Valdmanis}}, \ and\
  \bibinfo {author} {\bibfnamefont {D.~A.}\ \bibnamefont {Kleinman}},\
  }\href@noop {} {\bibfield  {journal} {\bibinfo  {journal} {Phys. Rev. Lett.}\
  }\textbf {\bibinfo {volume} {53}},\ \bibinfo {pages} {1555} (\bibinfo {year}
  {1984})}\BibitemShut {NoStop}%
\bibitem [{\citenamefont {Cheung}\ and\ \citenamefont
  {Auston}(1985)}]{Cheung1985}%
  \BibitemOpen
  \bibfield  {author} {\bibinfo {author} {\bibfnamefont {K.~P.}\ \bibnamefont
  {Cheung}}\ and\ \bibinfo {author} {\bibfnamefont {D.~H.}\ \bibnamefont
  {Auston}},\ }\href@noop {} {\bibfield  {journal} {\bibinfo  {journal} {Phys.
  Rev. Lett.}\ }\textbf {\bibinfo {volume} {55}},\ \bibinfo {pages} {2152}
  (\bibinfo {year} {1985})}\BibitemShut {NoStop}%
\bibitem [{\citenamefont {Hu}\ \emph {et~al.}(1990)\citenamefont {Hu},
  \citenamefont {Zhang}, \citenamefont {Auston},\ and\ \citenamefont
  {Smith}}]{Hu1990}%
  \BibitemOpen
  \bibfield  {author} {\bibinfo {author} {\bibfnamefont {B.~B.}\ \bibnamefont
  {Hu}}, \bibinfo {author} {\bibfnamefont {X.-C.}\ \bibnamefont {Zhang}},
  \bibinfo {author} {\bibfnamefont {D.~H.}\ \bibnamefont {Auston}}, \ and\
  \bibinfo {author} {\bibfnamefont {P.~R.}\ \bibnamefont {Smith}},\ }\href@noop
  {} {\bibfield  {journal} {\bibinfo  {journal} {Appl. Phys. Lett.}\ }\textbf
  {\bibinfo {volume} {56}},\ \bibinfo {pages} {506} (\bibinfo {year}
  {1990})}\BibitemShut {NoStop}%
\bibitem [{\citenamefont {Stevens}\ \emph {et~al.}(2001)\citenamefont
  {Stevens}, \citenamefont {Wahlstrand}, \citenamefont {Kuhl},\ and\
  \citenamefont {Merlin}}]{Stevens2001}%
  \BibitemOpen
  \bibfield  {author} {\bibinfo {author} {\bibfnamefont {T.~E.}\ \bibnamefont
  {Stevens}}, \bibinfo {author} {\bibfnamefont {J.~K.}\ \bibnamefont
  {Wahlstrand}}, \bibinfo {author} {\bibfnamefont {J.}~\bibnamefont {Kuhl}}, \
  and\ \bibinfo {author} {\bibfnamefont {R.}~\bibnamefont {Merlin}},\
  }\href@noop {} {\bibfield  {journal} {\bibinfo  {journal} {Science}\ }\textbf
  {\bibinfo {volume} {291}},\ \bibinfo {pages} {627} (\bibinfo {year}
  {2001})}\BibitemShut {NoStop}%
\bibitem [{\citenamefont {Hebling}\ \emph {et~al.}(2002)\citenamefont
  {Hebling}, \citenamefont {Alm\'{a}si}, \citenamefont {Kozma},\ and\
  \citenamefont {Kuhl}}]{Hebling2002}%
  \BibitemOpen
  \bibfield  {author} {\bibinfo {author} {\bibfnamefont {J.}~\bibnamefont
  {Hebling}}, \bibinfo {author} {\bibfnamefont {G.}~\bibnamefont {Alm\'{a}si}},
  \bibinfo {author} {\bibfnamefont {I.~Z.}\ \bibnamefont {Kozma}}, \ and\
  \bibinfo {author} {\bibfnamefont {J.}~\bibnamefont {Kuhl}},\ }\href@noop {}
  {\bibfield  {journal} {\bibinfo  {journal} {Opt. Express}\ }\textbf {\bibinfo
  {volume} {10}},\ \bibinfo {pages} {1161} (\bibinfo {year}
  {2002})}\BibitemShut {NoStop}%
\bibitem [{\citenamefont {Suizu}\ \emph {et~al.}(2009)\citenamefont {Suizu},
  \citenamefont {Koketsu}, \citenamefont {Shibuya}, \citenamefont {Tsutsui},
  \citenamefont {Akiba},\ and\ \citenamefont {Kawase}}]{Suizu2009}%
  \BibitemOpen
  \bibfield  {author} {\bibinfo {author} {\bibfnamefont {K.}~\bibnamefont
  {Suizu}}, \bibinfo {author} {\bibfnamefont {K.}~\bibnamefont {Koketsu}},
  \bibinfo {author} {\bibfnamefont {T.}~\bibnamefont {Shibuya}}, \bibinfo
  {author} {\bibfnamefont {T.}~\bibnamefont {Tsutsui}}, \bibinfo {author}
  {\bibfnamefont {T.}~\bibnamefont {Akiba}}, \ and\ \bibinfo {author}
  {\bibfnamefont {K.}~\bibnamefont {Kawase}},\ }\href {\doibase
  10.1364/OE.17.006676} {\bibfield  {journal} {\bibinfo  {journal} {Opt.
  Express}\ }\textbf {\bibinfo {volume} {17}},\ \bibinfo {pages} {6676}
  (\bibinfo {year} {2009})}\BibitemShut {NoStop}%
\bibitem [{\citenamefont {Crimmins}\ \emph {et~al.}(2002)\citenamefont
  {Crimmins}, \citenamefont {Stoyanov},\ and\ \citenamefont
  {Nelson}}]{Crimmins2002}%
  \BibitemOpen
  \bibfield  {author} {\bibinfo {author} {\bibfnamefont {T.~F.}\ \bibnamefont
  {Crimmins}}, \bibinfo {author} {\bibfnamefont {N.~S.}\ \bibnamefont
  {Stoyanov}}, \ and\ \bibinfo {author} {\bibfnamefont {K.~A.}\ \bibnamefont
  {Nelson}},\ }\href@noop {} {\bibfield  {journal} {\bibinfo  {journal} {J.
  Chem. Phys.}\ }\textbf {\bibinfo {volume} {117}},\ \bibinfo {pages} {2882}
  (\bibinfo {year} {2002})}\BibitemShut {NoStop}%
\bibitem [{\citenamefont {Stoyanov}\ \emph {et~al.}(2002)\citenamefont
  {Stoyanov}, \citenamefont {Ward}, \citenamefont {Feurer},\ and\ \citenamefont
  {Nelson}}]{Stoyanov2002}%
  \BibitemOpen
  \bibfield  {author} {\bibinfo {author} {\bibfnamefont {N.~S.}\ \bibnamefont
  {Stoyanov}}, \bibinfo {author} {\bibfnamefont {D.~W.}\ \bibnamefont {Ward}},
  \bibinfo {author} {\bibfnamefont {T.}~\bibnamefont {Feurer}}, \ and\ \bibinfo
  {author} {\bibfnamefont {K.~A.}\ \bibnamefont {Nelson}},\ }\href {\doibase
  10.1063/1.1491952} {\bibfield  {journal} {\bibinfo  {journal} {J. Chem.
  Phys.}\ }\textbf {\bibinfo {volume} {117}},\ \bibinfo {pages} {2897}
  (\bibinfo {year} {2002})}\BibitemShut {NoStop}%
\bibitem [{\citenamefont {Feurer}\ \emph {et~al.}(2007)\citenamefont {Feurer},
  \citenamefont {Stoyanov}, \citenamefont {Ward}, \citenamefont {Vaughan},
  \citenamefont {Statz},\ and\ \citenamefont {Nelson}}]{Feurer2007}%
  \BibitemOpen
  \bibfield  {author} {\bibinfo {author} {\bibfnamefont {T.}~\bibnamefont
  {Feurer}}, \bibinfo {author} {\bibfnamefont {N.~S.}\ \bibnamefont
  {Stoyanov}}, \bibinfo {author} {\bibfnamefont {D.~W.}\ \bibnamefont {Ward}},
  \bibinfo {author} {\bibfnamefont {J.~C.}\ \bibnamefont {Vaughan}}, \bibinfo
  {author} {\bibfnamefont {E.~R.}\ \bibnamefont {Statz}}, \ and\ \bibinfo
  {author} {\bibfnamefont {K.~A.}\ \bibnamefont {Nelson}},\ }\href {\doibase
  10.1146/annurev.matsci.37.052506.084327} {\bibfield  {journal} {\bibinfo
  {journal} {Annu. Rev. Mater. Res.}\ }\textbf {\bibinfo {volume} {37}},\
  \bibinfo {pages} {317} (\bibinfo {year} {2007})}\BibitemShut {NoStop}%
\bibitem [{\citenamefont {Yang}\ \emph {et~al.}(2010)\citenamefont {Yang},
  \citenamefont {Wu}, \citenamefont {Xu}, \citenamefont {Nelson},\ and\
  \citenamefont {Werley}}]{Yang2010}%
  \BibitemOpen
  \bibfield  {author} {\bibinfo {author} {\bibfnamefont {C.}~\bibnamefont
  {Yang}}, \bibinfo {author} {\bibfnamefont {Q.}~\bibnamefont {Wu}}, \bibinfo
  {author} {\bibfnamefont {J.}~\bibnamefont {Xu}}, \bibinfo {author}
  {\bibfnamefont {K.~A.}\ \bibnamefont {Nelson}}, \ and\ \bibinfo {author}
  {\bibfnamefont {C.~A.}\ \bibnamefont {Werley}},\ }\href@noop {} {\bibfield
  {journal} {\bibinfo  {journal} {Opt. Express}\ }\textbf {\bibinfo {volume}
  {18}},\ \bibinfo {pages} {26351} (\bibinfo {year} {2010})}\BibitemShut
  {NoStop}%
\bibitem [{\citenamefont {Wang}\ \emph {et~al.}(2015)\citenamefont {Wang},
  \citenamefont {Su},\ and\ \citenamefont {Hegmann}}]{Wang2015}%
  \BibitemOpen
  \bibfield  {author} {\bibinfo {author} {\bibfnamefont {Z.}~\bibnamefont
  {Wang}}, \bibinfo {author} {\bibfnamefont {F.}~\bibnamefont {Su}}, \ and\
  \bibinfo {author} {\bibfnamefont {F.~A.}\ \bibnamefont {Hegmann}},\
  }\href@noop {} {\bibfield  {journal} {\bibinfo  {journal} {Opt. Express}\
  }\textbf {\bibinfo {volume} {23}},\ \bibinfo {pages} {8073} (\bibinfo {year}
  {2015})}\BibitemShut {NoStop}%
\bibitem [{\citenamefont {Kawase}\ \emph {et~al.}(1996)\citenamefont {Kawase},
  \citenamefont {Sato}, \citenamefont {Taniuchi},\ and\ \citenamefont
  {Ito}}]{Kawase1996}%
  \BibitemOpen
  \bibfield  {author} {\bibinfo {author} {\bibfnamefont {K.}~\bibnamefont
  {Kawase}}, \bibinfo {author} {\bibfnamefont {M.}~\bibnamefont {Sato}},
  \bibinfo {author} {\bibfnamefont {T.}~\bibnamefont {Taniuchi}}, \ and\
  \bibinfo {author} {\bibfnamefont {H.}~\bibnamefont {Ito}},\ }\href {\doibase
  10.1063/1.115828} {\bibfield  {journal} {\bibinfo  {journal} {Appl. Phys.
  Lett.}\ }\textbf {\bibinfo {volume} {68}},\ \bibinfo {pages} {2483} (\bibinfo
  {year} {1996})}\BibitemShut {NoStop}%
\bibitem [{\citenamefont {Kawase}\ \emph {et~al.}(2002)\citenamefont {Kawase},
  \citenamefont {Shikata},\ and\ \citenamefont {Ito}}]{Kawase2002}%
  \BibitemOpen
  \bibfield  {author} {\bibinfo {author} {\bibfnamefont {K.}~\bibnamefont
  {Kawase}}, \bibinfo {author} {\bibfnamefont {J.}~\bibnamefont {Shikata}}, \
  and\ \bibinfo {author} {\bibfnamefont {H.}~\bibnamefont {Ito}},\ }\href
  {\doibase 10.1088/0022-3727/35/3/201} {\bibfield  {journal} {\bibinfo
  {journal} {J. Phys. D: Appl. Phys.}\ }\textbf {\bibinfo {volume} {35}},\
  \bibinfo {pages} {R1} (\bibinfo {year} {2002})}\BibitemShut {NoStop}%
\bibitem [{\citenamefont {Yeh}\ \emph {et~al.}(2007)\citenamefont {Yeh},
  \citenamefont {Hoffmann}, \citenamefont {Hebling},\ and\ \citenamefont
  {Nelson}}]{Yeh2007}%
  \BibitemOpen
  \bibfield  {author} {\bibinfo {author} {\bibfnamefont {K.-L.}\ \bibnamefont
  {Yeh}}, \bibinfo {author} {\bibfnamefont {M.~C.}\ \bibnamefont {Hoffmann}},
  \bibinfo {author} {\bibfnamefont {J.}~\bibnamefont {Hebling}}, \ and\
  \bibinfo {author} {\bibfnamefont {K.~A.}\ \bibnamefont {Nelson}},\ }\href
  {\doibase 10.1063/1.2734374} {\bibfield  {journal} {\bibinfo  {journal}
  {Appl. Phys. Lett.}\ }\textbf {\bibinfo {volume} {90}},\ \bibinfo {pages}
  {171121} (\bibinfo {year} {2007})}\BibitemShut {NoStop}%
\bibitem [{\citenamefont {{Fejer}}\ \emph {et~al.}(1992)\citenamefont
  {{Fejer}}, \citenamefont {{Magel}}, \citenamefont {{Jundt}},\ and\
  \citenamefont {{Byer}}}]{Fejer1992}%
  \BibitemOpen
  \bibfield  {author} {\bibinfo {author} {\bibfnamefont {M.~M.}\ \bibnamefont
  {{Fejer}}}, \bibinfo {author} {\bibfnamefont {G.~A.}\ \bibnamefont
  {{Magel}}}, \bibinfo {author} {\bibfnamefont {D.~H.}\ \bibnamefont
  {{Jundt}}}, \ and\ \bibinfo {author} {\bibfnamefont {R.~L.}\ \bibnamefont
  {{Byer}}},\ }\href@noop {} {\bibfield  {journal} {\bibinfo  {journal} {IEEE
  J. Quantum Electron.}\ }\textbf {\bibinfo {volume} {28}},\ \bibinfo {pages}
  {2631} (\bibinfo {year} {1992})}\BibitemShut {NoStop}%
\bibitem [{\citenamefont {Yamada}\ \emph {et~al.}(1993)\citenamefont {Yamada},
  \citenamefont {Nada}, \citenamefont {Saitoh},\ and\ \citenamefont
  {Watanabe}}]{Yamada1993}%
  \BibitemOpen
  \bibfield  {author} {\bibinfo {author} {\bibfnamefont {M.}~\bibnamefont
  {Yamada}}, \bibinfo {author} {\bibfnamefont {N.}~\bibnamefont {Nada}},
  \bibinfo {author} {\bibfnamefont {M.}~\bibnamefont {Saitoh}}, \ and\ \bibinfo
  {author} {\bibfnamefont {K.}~\bibnamefont {Watanabe}},\ }\href {\doibase
  10.1063/1.108925} {\bibfield  {journal} {\bibinfo  {journal} {Appl. Phys.
  Lett.}\ }\textbf {\bibinfo {volume} {62}},\ \bibinfo {pages} {435} (\bibinfo
  {year} {1993})}\BibitemShut {NoStop}%
\bibitem [{\citenamefont {{Burns}}\ \emph {et~al.}(1994)\citenamefont
  {{Burns}}, \citenamefont {{McElhanon}},\ and\ \citenamefont
  {{Goldberg}}}]{Burns1994}%
  \BibitemOpen
  \bibfield  {author} {\bibinfo {author} {\bibfnamefont {W.~K.}\ \bibnamefont
  {{Burns}}}, \bibinfo {author} {\bibfnamefont {W.}~\bibnamefont
  {{McElhanon}}}, \ and\ \bibinfo {author} {\bibfnamefont {L.}~\bibnamefont
  {{Goldberg}}},\ }\href@noop {} {\bibfield  {journal} {\bibinfo  {journal}
  {IEEE Photon. Technol. Lett.}\ }\textbf {\bibinfo {volume} {6}},\ \bibinfo
  {pages} {252} (\bibinfo {year} {1994})}\BibitemShut {NoStop}%
\bibitem [{\citenamefont {Myers}\ \emph {et~al.}(1995)\citenamefont {Myers},
  \citenamefont {Eckardt}, \citenamefont {Fejer}, \citenamefont {Byer},
  \citenamefont {Bosenberg},\ and\ \citenamefont {Pierce}}]{Myers1995}%
  \BibitemOpen
  \bibfield  {author} {\bibinfo {author} {\bibfnamefont {L.~E.}\ \bibnamefont
  {Myers}}, \bibinfo {author} {\bibfnamefont {R.~C.}\ \bibnamefont {Eckardt}},
  \bibinfo {author} {\bibfnamefont {M.~M.}\ \bibnamefont {Fejer}}, \bibinfo
  {author} {\bibfnamefont {R.~L.}\ \bibnamefont {Byer}}, \bibinfo {author}
  {\bibfnamefont {W.~R.}\ \bibnamefont {Bosenberg}}, \ and\ \bibinfo {author}
  {\bibfnamefont {J.~W.}\ \bibnamefont {Pierce}},\ }\href@noop {} {\bibfield
  {journal} {\bibinfo  {journal} {J. Opt. Soc. Am. B}\ }\textbf {\bibinfo
  {volume} {12}},\ \bibinfo {pages} {2102} (\bibinfo {year}
  {1995})}\BibitemShut {NoStop}%
\bibitem [{\citenamefont {Stepanov}\ \emph {et~al.}(2003)\citenamefont
  {Stepanov}, \citenamefont {Hebling},\ and\ \citenamefont
  {Kuhl}}]{Stepanov2003}%
  \BibitemOpen
  \bibfield  {author} {\bibinfo {author} {\bibfnamefont {A.~G.}\ \bibnamefont
  {Stepanov}}, \bibinfo {author} {\bibfnamefont {J.}~\bibnamefont {Hebling}}, \
  and\ \bibinfo {author} {\bibfnamefont {J.}~\bibnamefont {Kuhl}},\ }\href@noop
  {} {\bibfield  {journal} {\bibinfo  {journal} {Appl. Phys. Lett.}\ }\textbf
  {\bibinfo {volume} {83}},\ \bibinfo {pages} {3000} (\bibinfo {year}
  {2003})}\BibitemShut {NoStop}%
\bibitem [{\citenamefont {Hebling}\ \emph {et~al.}(2004)\citenamefont
  {Hebling}, \citenamefont {Stepanov}, \citenamefont {Alm{\'a}si},
  \citenamefont {Bartal},\ and\ \citenamefont {Kuhl}}]{Hebling2004}%
  \BibitemOpen
  \bibfield  {author} {\bibinfo {author} {\bibfnamefont {J.}~\bibnamefont
  {Hebling}}, \bibinfo {author} {\bibfnamefont {A.~G.}\ \bibnamefont
  {Stepanov}}, \bibinfo {author} {\bibfnamefont {G.}~\bibnamefont
  {Alm{\'a}si}}, \bibinfo {author} {\bibfnamefont {B.}~\bibnamefont {Bartal}},
  \ and\ \bibinfo {author} {\bibfnamefont {J.}~\bibnamefont {Kuhl}},\
  }\href@noop {} {\bibfield  {journal} {\bibinfo  {journal} {Appl. Phys. B}\
  }\textbf {\bibinfo {volume} {78}},\ \bibinfo {pages} {593} (\bibinfo {year}
  {2004})}\BibitemShut {NoStop}%
\bibitem [{\citenamefont {Stepanov}\ \emph {et~al.}(2005)\citenamefont
  {Stepanov}, \citenamefont {Kuhl}, \citenamefont {Kozma}, \citenamefont
  {Riedle}, \citenamefont {Alm\'{a}si},\ and\ \citenamefont
  {Hebling}}]{Stepanov2005}%
  \BibitemOpen
  \bibfield  {author} {\bibinfo {author} {\bibfnamefont {A.~G.}\ \bibnamefont
  {Stepanov}}, \bibinfo {author} {\bibfnamefont {J.}~\bibnamefont {Kuhl}},
  \bibinfo {author} {\bibfnamefont {I.~Z.}\ \bibnamefont {Kozma}}, \bibinfo
  {author} {\bibfnamefont {E.}~\bibnamefont {Riedle}}, \bibinfo {author}
  {\bibfnamefont {G.}~\bibnamefont {Alm\'{a}si}}, \ and\ \bibinfo {author}
  {\bibfnamefont {J.}~\bibnamefont {Hebling}},\ }\href@noop {} {\bibfield
  {journal} {\bibinfo  {journal} {Opt. Express}\ }\textbf {\bibinfo {volume}
  {13}},\ \bibinfo {pages} {5762} (\bibinfo {year} {2005})}\BibitemShut
  {NoStop}%
\bibitem [{\citenamefont {Hirori}\ \emph {et~al.}(2011)\citenamefont {Hirori},
  \citenamefont {Doi}, \citenamefont {Blanchard},\ and\ \citenamefont
  {Tanaka}}]{Hirori2011}%
  \BibitemOpen
  \bibfield  {author} {\bibinfo {author} {\bibfnamefont {H.}~\bibnamefont
  {Hirori}}, \bibinfo {author} {\bibfnamefont {A.}~\bibnamefont {Doi}},
  \bibinfo {author} {\bibfnamefont {F.}~\bibnamefont {Blanchard}}, \ and\
  \bibinfo {author} {\bibfnamefont {K.}~\bibnamefont {Tanaka}},\ }\href@noop {}
  {\bibfield  {journal} {\bibinfo  {journal} {Appl. Phys. Lett.}\ }\textbf
  {\bibinfo {volume} {98}},\ \bibinfo {pages} {091106} (\bibinfo {year}
  {2011})}\BibitemShut {NoStop}%
\bibitem [{\citenamefont {Nagai}\ \emph {et~al.}(2012)\citenamefont {Nagai},
  \citenamefont {Matsubara},\ and\ \citenamefont {Ashida}}]{Nagai2012}%
  \BibitemOpen
  \bibfield  {author} {\bibinfo {author} {\bibfnamefont {M.}~\bibnamefont
  {Nagai}}, \bibinfo {author} {\bibfnamefont {E.}~\bibnamefont {Matsubara}}, \
  and\ \bibinfo {author} {\bibfnamefont {M.}~\bibnamefont {Ashida}},\
  }\href@noop {} {\bibfield  {journal} {\bibinfo  {journal} {Opt. Express}\
  }\textbf {\bibinfo {volume} {20}},\ \bibinfo {pages} {6509} (\bibinfo {year}
  {2012})}\BibitemShut {NoStop}%
\bibitem [{\citenamefont {Kawase}\ \emph {et~al.}(2001)\citenamefont {Kawase},
  \citenamefont {Shikata}, \citenamefont {Minamide}, \citenamefont {Imai},\
  and\ \citenamefont {Ito}}]{Kawase2001}%
  \BibitemOpen
  \bibfield  {author} {\bibinfo {author} {\bibfnamefont {K.}~\bibnamefont
  {Kawase}}, \bibinfo {author} {\bibfnamefont {J.}~\bibnamefont {Shikata}},
  \bibinfo {author} {\bibfnamefont {H.}~\bibnamefont {Minamide}}, \bibinfo
  {author} {\bibfnamefont {K.}~\bibnamefont {Imai}}, \ and\ \bibinfo {author}
  {\bibfnamefont {H.}~\bibnamefont {Ito}},\ }\href@noop {} {\bibfield
  {journal} {\bibinfo  {journal} {Appl. Opt.}\ }\textbf {\bibinfo {volume}
  {40}},\ \bibinfo {pages} {1423} (\bibinfo {year} {2001})}\BibitemShut
  {NoStop}%
\bibitem [{\citenamefont {Bodrov}\ \emph {et~al.}(2008)\citenamefont {Bodrov},
  \citenamefont {Bakunov},\ and\ \citenamefont {Hangyo}}]{Bodrov2008}%
  \BibitemOpen
  \bibfield  {author} {\bibinfo {author} {\bibfnamefont {S.~B.}\ \bibnamefont
  {Bodrov}}, \bibinfo {author} {\bibfnamefont {M.~I.}\ \bibnamefont {Bakunov}},
  \ and\ \bibinfo {author} {\bibfnamefont {M.}~\bibnamefont {Hangyo}},\ }\href
  {\doibase 10.1063/1.3005987} {\bibfield  {journal} {\bibinfo  {journal} {J.
  Appl. Phys.}\ }\textbf {\bibinfo {volume} {104}},\ \bibinfo {pages} {093105}
  (\bibinfo {year} {2008})}\BibitemShut {NoStop}%
\bibitem [{\citenamefont {Tani}\ \emph {et~al.}(2011)\citenamefont {Tani},
  \citenamefont {Horita}, \citenamefont {Kinoshita}, \citenamefont {Que},
  \citenamefont {Estacio}, \citenamefont {Yamamoto},\ and\ \citenamefont
  {Bakunov}}]{Tani2011}%
  \BibitemOpen
  \bibfield  {author} {\bibinfo {author} {\bibfnamefont {M.}~\bibnamefont
  {Tani}}, \bibinfo {author} {\bibfnamefont {K.}~\bibnamefont {Horita}},
  \bibinfo {author} {\bibfnamefont {T.}~\bibnamefont {Kinoshita}}, \bibinfo
  {author} {\bibfnamefont {C.~T.}\ \bibnamefont {Que}}, \bibinfo {author}
  {\bibfnamefont {E.}~\bibnamefont {Estacio}}, \bibinfo {author} {\bibfnamefont
  {K.}~\bibnamefont {Yamamoto}}, \ and\ \bibinfo {author} {\bibfnamefont
  {M.~I.}\ \bibnamefont {Bakunov}},\ }\href@noop {} {\bibfield  {journal}
  {\bibinfo  {journal} {Opt. Express}\ }\textbf {\bibinfo {volume} {19}},\
  \bibinfo {pages} {19901} (\bibinfo {year} {2011})}\BibitemShut {NoStop}%
\bibitem [{\citenamefont {Bakunov}\ \emph {et~al.}(2012)\citenamefont
  {Bakunov}, \citenamefont {Mashkovich}, \citenamefont {Tsarev},\ and\
  \citenamefont {Gorelov}}]{Bakunov2012}%
  \BibitemOpen
  \bibfield  {author} {\bibinfo {author} {\bibfnamefont {M.~I.}\ \bibnamefont
  {Bakunov}}, \bibinfo {author} {\bibfnamefont {E.~A.}\ \bibnamefont
  {Mashkovich}}, \bibinfo {author} {\bibfnamefont {M.~V.}\ \bibnamefont
  {Tsarev}}, \ and\ \bibinfo {author} {\bibfnamefont {S.~D.}\ \bibnamefont
  {Gorelov}},\ }\href@noop {} {\bibfield  {journal} {\bibinfo  {journal} {Appl.
  Phys. Lett.}\ }\textbf {\bibinfo {volume} {101}},\ \bibinfo {pages} {151102}
  (\bibinfo {year} {2012})}\BibitemShut {NoStop}%
\bibitem [{\citenamefont {Bakunov}\ \emph {et~al.}(2014)\citenamefont
  {Bakunov}, \citenamefont {Mashkovich},\ and\ \citenamefont
  {Svinkina}}]{Bakunov2014}%
  \BibitemOpen
  \bibfield  {author} {\bibinfo {author} {\bibfnamefont {M.~I.}\ \bibnamefont
  {Bakunov}}, \bibinfo {author} {\bibfnamefont {E.~A.}\ \bibnamefont
  {Mashkovich}}, \ and\ \bibinfo {author} {\bibfnamefont {E.~V.}\ \bibnamefont
  {Svinkina}},\ }\href {\doibase 10.1364/OL.39.006779} {\bibfield  {journal}
  {\bibinfo  {journal} {Opt. Lett.}\ }\textbf {\bibinfo {volume} {39}},\
  \bibinfo {pages} {6779} (\bibinfo {year} {2014})}\BibitemShut {NoStop}%
\bibitem [{\citenamefont {Satoh}\ \emph {et~al.}(2015)\citenamefont {Satoh},
  \citenamefont {Iida}, \citenamefont {Higuchi}, \citenamefont {Fiebig},\ and\
  \citenamefont {Shimura}}]{Satoh2015}%
  \BibitemOpen
  \bibfield  {author} {\bibinfo {author} {\bibfnamefont {T.}~\bibnamefont
  {Satoh}}, \bibinfo {author} {\bibfnamefont {R.}~\bibnamefont {Iida}},
  \bibinfo {author} {\bibfnamefont {T.}~\bibnamefont {Higuchi}}, \bibinfo
  {author} {\bibfnamefont {M.}~\bibnamefont {Fiebig}}, \ and\ \bibinfo {author}
  {\bibfnamefont {T.}~\bibnamefont {Shimura}},\ }\href {\doibase
  10.1038/nphoton.2014.273} {\bibfield  {journal} {\bibinfo  {journal} {Nat.
  Photon.}\ }\textbf {\bibinfo {volume} {9}},\ \bibinfo {pages} {25} (\bibinfo
  {year} {2015})}\BibitemShut {NoStop}%
\bibitem [{\citenamefont {Yoshimine}\ \emph {et~al.}(2014)\citenamefont
  {Yoshimine}, \citenamefont {Satoh}, \citenamefont {Iida}, \citenamefont
  {Stupakiewicz}, \citenamefont {Maziewski},\ and\ \citenamefont
  {Shimura}}]{Yoshimine2014}%
  \BibitemOpen
  \bibfield  {author} {\bibinfo {author} {\bibfnamefont {I.}~\bibnamefont
  {Yoshimine}}, \bibinfo {author} {\bibfnamefont {T.}~\bibnamefont {Satoh}},
  \bibinfo {author} {\bibfnamefont {R.}~\bibnamefont {Iida}}, \bibinfo {author}
  {\bibfnamefont {A.}~\bibnamefont {Stupakiewicz}}, \bibinfo {author}
  {\bibfnamefont {A.}~\bibnamefont {Maziewski}}, \ and\ \bibinfo {author}
  {\bibfnamefont {T.}~\bibnamefont {Shimura}},\ }\href@noop {} {\bibfield
  {journal} {\bibinfo  {journal} {J. Appl. Phys.}\ }\textbf {\bibinfo {volume}
  {116}},\ \bibinfo {pages} {043907} (\bibinfo {year} {2014})}\BibitemShut
  {NoStop}%
\bibitem [{\citenamefont {Zelmon}\ \emph {et~al.}(1997)\citenamefont {Zelmon},
  \citenamefont {Small},\ and\ \citenamefont {Jundt}}]{Zelmon1997}%
  \BibitemOpen
  \bibfield  {author} {\bibinfo {author} {\bibfnamefont {D.~E.}\ \bibnamefont
  {Zelmon}}, \bibinfo {author} {\bibfnamefont {D.~L.}\ \bibnamefont {Small}}, \
  and\ \bibinfo {author} {\bibfnamefont {D.}~\bibnamefont {Jundt}},\ }\href
  {\doibase 10.1364/JOSAB.14.003319} {\bibfield  {journal} {\bibinfo  {journal}
  {J. Opt. Soc. Am. B}\ }\textbf {\bibinfo {volume} {14}},\ \bibinfo {pages}
  {3319} (\bibinfo {year} {1997})}\BibitemShut {NoStop}%
\bibitem [{\citenamefont {Loudon}(1964)}]{Loudon1964}%
  \BibitemOpen
  \bibfield  {author} {\bibinfo {author} {\bibfnamefont {R.}~\bibnamefont
  {Loudon}},\ }\href@noop {} {\bibfield  {journal} {\bibinfo  {journal} {Adv.
  Phys.}\ }\textbf {\bibinfo {volume} {13}},\ \bibinfo {pages} {423} (\bibinfo
  {year} {1964})}\BibitemShut {NoStop}%
\bibitem [{\citenamefont {Bakunov}\ \emph {et~al.}(2005)\citenamefont
  {Bakunov}, \citenamefont {Maslov},\ and\ \citenamefont
  {Bodrov}}]{Bakunov2005}%
  \BibitemOpen
  \bibfield  {author} {\bibinfo {author} {\bibfnamefont {M.~I.}\ \bibnamefont
  {Bakunov}}, \bibinfo {author} {\bibfnamefont {A.~V.}\ \bibnamefont {Maslov}},
  \ and\ \bibinfo {author} {\bibfnamefont {S.~B.}\ \bibnamefont {Bodrov}},\
  }\href {\doibase 10.1103/PhysRevB.72.195336} {\bibfield  {journal} {\bibinfo
  {journal} {Phys. Rev. B}\ }\textbf {\bibinfo {volume} {72}},\ \bibinfo
  {pages} {195336} (\bibinfo {year} {2005})}\BibitemShut {NoStop}%
\bibitem [{\citenamefont {Shen}(2002)}]{Shen_nonlinearOptics}%
  \BibitemOpen
  \bibfield  {author} {\bibinfo {author} {\bibfnamefont {Y.~R.}\ \bibnamefont
  {Shen}},\ }\href@noop {} {\emph {\bibinfo {title} {The Principles of
  Nonlinear Optics}}}\ (\bibinfo  {publisher} {Wiley, New York},\ \bibinfo
  {year} {2002})\BibitemShut {NoStop}%
\bibitem [{\citenamefont {Khan}\ \emph {et~al.}(2020)\citenamefont {Khan},
  \citenamefont {Kanamaru}, \citenamefont {Matsumoto}, \citenamefont {Ito},\
  and\ \citenamefont {Satoh}}]{Khan2020}%
  \BibitemOpen
  \bibfield  {author} {\bibinfo {author} {\bibfnamefont {P.}~\bibnamefont
  {Khan}}, \bibinfo {author} {\bibfnamefont {M.}~\bibnamefont {Kanamaru}},
  \bibinfo {author} {\bibfnamefont {K.}~\bibnamefont {Matsumoto}}, \bibinfo
  {author} {\bibfnamefont {T.}~\bibnamefont {Ito}}, \ and\ \bibinfo {author}
  {\bibfnamefont {T.}~\bibnamefont {Satoh}},\ }\href@noop {} {\bibfield
  {journal} {\bibinfo  {journal} {Phys. Rev. B}\ }\textbf {\bibinfo {volume}
  {101}},\ \bibinfo {pages} {134413} (\bibinfo {year} {2020})}\BibitemShut
  {NoStop}%
\bibitem [{SM2()}]{SM20}%
  \BibitemOpen
  \href@noop {} {}\bibinfo {note} {See Supplemental
  Movies\\http://link.aps.org/supplemental/10.1103/PhysRevB.102.094313 for the
  experimentally observed phonon-polariton and numerically calculated THz
  electric field. Note that the accumulation of Fig. 2 is different from that
  of the movies.}\BibitemShut {Stop}%
\bibitem [{\citenamefont {Claus}(1972)}]{Claus1972}%
  \BibitemOpen
  \bibfield  {author} {\bibinfo {author} {\bibfnamefont {R.}~\bibnamefont
  {Claus}},\ }\href@noop {} {\bibfield  {journal} {\bibinfo  {journal} {Phys.
  Status Solidi B}\ }\textbf {\bibinfo {volume} {50}},\ \bibinfo {pages} {11}
  (\bibinfo {year} {1972})}\BibitemShut {NoStop}%
\bibitem [{\citenamefont {Delbart}\ \emph {et~al.}(1998)\citenamefont
  {Delbart}, \citenamefont {Derr{\'e}},\ and\ \citenamefont
  {Chipaux}}]{Delbart1998}%
  \BibitemOpen
  \bibfield  {author} {\bibinfo {author} {\bibfnamefont {A.}~\bibnamefont
  {Delbart}}, \bibinfo {author} {\bibfnamefont {J.}~\bibnamefont {Derr{\'e}}},
  \ and\ \bibinfo {author} {\bibfnamefont {R.}~\bibnamefont {Chipaux}},\
  }\href@noop {} {\bibfield  {journal} {\bibinfo  {journal} {Eur. Phys. J. D}\
  }\textbf {\bibinfo {volume} {1}},\ \bibinfo {pages} {109} (\bibinfo {year}
  {1998})}\BibitemShut {NoStop}%
\bibitem [{\citenamefont {Barker}\ and\ \citenamefont
  {Loudon}(1967)}]{Barker1967}%
  \BibitemOpen
  \bibfield  {author} {\bibinfo {author} {\bibfnamefont {A.~S.}\ \bibnamefont
  {Barker}}\ and\ \bibinfo {author} {\bibfnamefont {R.}~\bibnamefont
  {Loudon}},\ }\href@noop {} {\bibfield  {journal} {\bibinfo  {journal} {Phys.
  Rev.}\ }\textbf {\bibinfo {volume} {158}},\ \bibinfo {pages} {433} (\bibinfo
  {year} {1967})}\BibitemShut {NoStop}%
\bibitem [{\citenamefont {Satoh}\ \emph {et~al.}(2012)\citenamefont {Satoh},
  \citenamefont {Terui}, \citenamefont {Moriya}, \citenamefont {Ivanov},
  \citenamefont {Ando}, \citenamefont {Saitoh}, \citenamefont {Shimura},\ and\
  \citenamefont {Kuroda}}]{Satoh2012}%
  \BibitemOpen
  \bibfield  {author} {\bibinfo {author} {\bibfnamefont {T.}~\bibnamefont
  {Satoh}}, \bibinfo {author} {\bibfnamefont {Y.}~\bibnamefont {Terui}},
  \bibinfo {author} {\bibfnamefont {R.}~\bibnamefont {Moriya}}, \bibinfo
  {author} {\bibfnamefont {B.~A.}\ \bibnamefont {Ivanov}}, \bibinfo {author}
  {\bibfnamefont {K.}~\bibnamefont {Ando}}, \bibinfo {author} {\bibfnamefont
  {E.}~\bibnamefont {Saitoh}}, \bibinfo {author} {\bibfnamefont
  {T.}~\bibnamefont {Shimura}}, \ and\ \bibinfo {author} {\bibfnamefont
  {K.}~\bibnamefont {Kuroda}},\ }\href@noop {} {\bibfield  {journal} {\bibinfo
  {journal} {Nat. Photon.}\ }\textbf {\bibinfo {volume} {6}},\ \bibinfo {pages}
  {662} (\bibinfo {year} {2012})}\BibitemShut {NoStop}%
\bibitem [{\citenamefont {Yoshimine}\ \emph {et~al.}(2017)\citenamefont
  {Yoshimine}, \citenamefont {Tanaka}, \citenamefont {Shimura},\ and\
  \citenamefont {Satoh}}]{Yoshimine2017}%
  \BibitemOpen
  \bibfield  {author} {\bibinfo {author} {\bibfnamefont {I.}~\bibnamefont
  {Yoshimine}}, \bibinfo {author} {\bibfnamefont {Y.~Y.}\ \bibnamefont
  {Tanaka}}, \bibinfo {author} {\bibfnamefont {T.}~\bibnamefont {Shimura}}, \
  and\ \bibinfo {author} {\bibfnamefont {T.}~\bibnamefont {Satoh}},\
  }\href@noop {} {\bibfield  {journal} {\bibinfo  {journal} {Europhys. Lett.}\
  }\textbf {\bibinfo {volume} {117}},\ \bibinfo {pages} {67001} (\bibinfo
  {year} {2017})}\BibitemShut {NoStop}%
\bibitem [{\citenamefont {van Tilburg}\ \emph {et~al.}(2017)\citenamefont {van
  Tilburg}, \citenamefont {Buijnsters}, \citenamefont {Fasolino}, \citenamefont
  {{T. Rasing}},\ and\ \citenamefont {Katsnelson}}]{Tilburg2017}%
  \BibitemOpen
  \bibfield  {author} {\bibinfo {author} {\bibfnamefont {L.~J.~A.}\
  \bibnamefont {van Tilburg}}, \bibinfo {author} {\bibfnamefont {F.~J.}\
  \bibnamefont {Buijnsters}}, \bibinfo {author} {\bibfnamefont
  {A.}~\bibnamefont {Fasolino}}, \bibinfo {author} {\bibnamefont {{T.
  Rasing}}}, \ and\ \bibinfo {author} {\bibfnamefont {M.~I.}\ \bibnamefont
  {Katsnelson}},\ }\href@noop {} {\bibfield  {journal} {\bibinfo  {journal}
  {Phys. Rev. B}\ }\textbf {\bibinfo {volume} {96}},\ \bibinfo {pages} {054437}
  (\bibinfo {year} {2017})}\BibitemShut {NoStop}%
\bibitem [{\citenamefont {Matsumoto}\ \emph {et~al.}(2020)\citenamefont
  {Matsumoto}, \citenamefont {Yoshimine}, \citenamefont {Himeno}, \citenamefont
  {Shimura},\ and\ \citenamefont {Satoh}}]{Kmatsumoto2020}%
  \BibitemOpen
  \bibfield  {author} {\bibinfo {author} {\bibfnamefont {K.}~\bibnamefont
  {Matsumoto}}, \bibinfo {author} {\bibfnamefont {I.}~\bibnamefont
  {Yoshimine}}, \bibinfo {author} {\bibfnamefont {K.}~\bibnamefont {Himeno}},
  \bibinfo {author} {\bibfnamefont {T.}~\bibnamefont {Shimura}}, \ and\
  \bibinfo {author} {\bibfnamefont {T.}~\bibnamefont {Satoh}},\ }\href@noop {}
  {\bibfield  {journal} {\bibinfo  {journal} {Phys. Rev. B}\ }\textbf {\bibinfo
  {volume} {101}},\ \bibinfo {pages} {184407} (\bibinfo {year}
  {2020})}\BibitemShut {NoStop}%
\end{thebibliography}%
\addcontentsline{toc}{chapter}{\bibname}
\end{document}